\documentclass[fleqn,usenatbib]{mnras}

\usepackage[T1]{fontenc}

\DeclareRobustCommand{\VAN}[3]{#2}
\let\VANthebibliography\thebibliography
\def\thebibliography{\DeclareRobustCommand{\VAN}[3]{##3}\VANthebibliography}

\usepackage{graphicx}	% Including figure files
\usepackage{amsmath}	% Advanced maths commands
\usepackage{amssymb}	% Extra maths symbols

\usepackage{newtxtext,newtxmath}

\title[Molecular contrails ]{Molecular contrails -- triggered contraction by passages of massive  objects through molecular clouds}

\author[Li \& Shi]{
Guang-Xing Li,$^{1}$ \thanks{E-mail: gxli@ynu.edu.cn}
Xun Shi, $^{1}$ \thanks{E-mail: xun@ynu.edu.cn}
\\
$^{1}$South-Western Institute for Astronomy Research, Yunnan University, Kunming, 650500 Yunnan, P.R. China
}

\date{Accepted XXX. Received YYY; in original form ZZZ}

\pubyear{2020}

\begin{document}
\label{firstpage}
\pagerange{\pageref{firstpage}--\pageref{lastpage}}
\maketitle

\begin{abstract}
	We study the effects of passages of compact objects such as stars, star clusters and black holes through molecular clouds, and propose that 
	the gravitational interaction between the compact object and the ambient gas can lead to 
	the formation of thin and collimated features made of dense gas, which we call ``molecular contrails''. 
	  Supercritical contrails can collapse further leading to  triggered star formation.  
	  The width of a molecular contrail is determined by the mass and velocity of the compact object and the velocity dispersion of the ambient molecular medium. Under typical conditions in the Milky Way,
	 passages of stellar-mass objects lead to the formation of  width $d\gtrsim 0.01\;\rm {parsec}$ contrails, and passages of star clusters lead to the formation of $d\gtrsim 1\;\rm {parsec}$  contrails. 
	 We present a few molecular contrail candidates from both categories identified from ALMA 1.3mm continuum observations of star-forming regions and the $^{13}$CO(1-0) map from the Galactic Ring Survey respectively.
	  The contrails represent an overlooked channel where stars and gas in the Galactic
    disk interact to
	  form
	  structures. They also present a potential way of detecting dark compact objects in the Milky Way.
\end{abstract}


%%%%%%%%%%%%%%%%%%%%%%%%%%%%%%%%%%%%%%%%%%%%%%%%%%



\section{Introduction}

Molecular clouds constitute an inseparable part of the interstellar medium, and the collapse of molecular clouds is the very process that %is linked
leads to the formation of stars in the universe. Researches  \citep[e.g.][ and references therein]{2014prpl.conf....3D,cite-key} have shown that the evolution of a molecular
cloud is a complex process characterized by the interplay
among turbulence, gravity, magnetic fields, stellar feedback,
and galactic shear. However, gravity is the ultimate driver of molecular cloud collapse and star formation, on which we focus in this paper.

Star formation is mostly a result of molecular cloud collapse from its self-gravity.  {  However, this process can be modulated by the global gravitational field in a galaxy contributed by other components e.g. stars. }For example, spiral arms and bars in disk galaxies, believed to be caused by instabilities of a disk of stars \citep[][and references therein]{2016ARA&A..54..667S}, play important roles in triggering star formation \citep{1969ApJ...158..123R,2020MNRAS.491.2162P}. Also, in the center of galaxies such as the Milky Way, gravitational field from the {  stellar bulge} may lead to a strong shear which suppresses star formation \citep{2020ApJ...897...89L}. {  
During the late stage of star formation, gravity of the stars can also influence the evolution of the circumstellar disk via gravity during stellar fly-bys \citep{1993MNRAS.261..190C,2018MNRAS.475.2314W}.}

% Gravity is the ultimate driver of star formation, and the collapse of molecular
% cloud is mostly driven by gravity from the gas itself. However, the Milky Way is a
% complex systems made of stars, gas as well as dark matter, where collapse can be caused by a combination of factors. Stars can act collectively to trigger star formation. Spiral arms in
% disk galaxies are believed to be caused by instabilities of a disk of stars
% \citep[][and references therein]{2016ARA&A..54..667S}, and spiral arms and bars 
% play important roles in triggering star formation \citep{1969ApJ...158..123R,2020MNRAS.491.2162P}.
% In the center of Galaxies such as the Milky Way, the presence of stars may lead
% to strong shear which suppresses the star formations \citep{2020ApJ...897...89L}. During the late stage of star formation, stellar fly-bys can also influence the evolution of the circumstellar disk via gravity \citep{1993MNRAS.261..190C,2018MNRAS.475.2314W}.

Can gravity from individual stars, star clusters and other compact objects such as black holes affect star formation? In this paper, we propose that  gravity from  compact objects can initiate wakes in the surrounding
 molecular gas via gravitational interaction during their passages, which could lead to the formation of thin and collimated features which we call the “molecular contrails”.
Due to the prevalence of the turbulent motions, interstellar molecular gas is expected to appear chaotic as it usually does. On top of this, the contrails should stand out as straight concentrations of molecular gas of various lengths and widths. 

We expect that the contrails with widths $d\gtrsim 1 \;\rm pc$  can be  identified from surveys of Milky Way disk such as the Galactic Ring Survey \citep{2006ApJS..163..145J} where molecular gas is traced at  $\approx 40''$ resolution (which corresponds to a size of 1.6 pc at a distance of 8 kpc).  Contrails with width  $d\gtrsim 0.01 \;\rm pc$ can be identified from interferometer observations of dense {molecular} clumps where clustered star formation is taking place. By inspecting existing observational data, we are able to identify some contrail candidates. 

%in detail.
% \xun{ \sout{Using observational data, we demonstrate that these contrails are ubiquitous and they exist in different sizes.}}

\section{Contrail Formation}
We consider the passage of a compact object of mass $m_{*}$ through the molecular medium. We are interested in the fast passages where  the relative speed  $v_{*}$ is much larger than the typical velocity dispersion of the medium $\sigma_{\rm v, medium}$, and study the response of the medium. 

\subsection{Formation Mechanism}

At a distance $r$ from the trajectory, the velocity injected due to gravity from the compact object into the medium %measured in terms of velocity
is
\begin{equation}\label{eq:p}
  v_{\rm inject} \approx a \times t_{\rm inject} = G m_* / r^2  \times r / v_* \approx G m_* /
  (r v_*)\;. \end{equation}
In order to trigger contraction or collapse, we require that the momentum injection measured in terms of $v_{\rm inject} $ to be comparable to or larger than characteristic velocity of the medium $\sigma_{\rm medium}$,
\begin{equation}\label{eq:v}
  v_{\rm inject}  \gtrsim  \sigma_{\rm medium}\;. \end{equation}
%Combining Eqs. \ref{eq:p} and \ref{eq:v} yields
From these we obtain an analytical estimate of the radius within which the passage has an effect 
\begin{equation}\label{eq:d}
  r_{\rm wake} \approx  \frac{G m_*}{v_*
    \sigma_{\rm medium}}\;.
\end{equation}

Note that here we are focusing on the effect of momentum injection. There are other ways through which the object and the medium can interact. For example, during the passages, the compact objects might accrete matter from the medium.
We note that in our case where $v_* > \sigma_{\rm medium}$, the Bondi radius -- radius within which gas is expected to accrete onto the compact object is $r_{\rm Bondi} \approx G m_{*} / v_*^2$ \citep{1941MNRAS.101..227H,1952MNRAS.112..195B}. Thus $r_{\rm wake} / r_{\rm Bondi} \approx v_* / \sigma_{\rm medium}$, %c_{\rm s}$,
and the mass accretion rate scales as $\dot M \propto (v_*/ \sigma_{\rm medium})^{-3}$. % c_{\rm s})^{-3}$.
For a compact object with a large velocity $v_* \gg \sigma_{\rm medium}$, we have $r_{\rm wake} \gg r_{\rm Bondi}$, and the mass accretion from the medium is inefficient, i.e. the wake is the dominant effect compared to accretion.
We also note that during the passage, a wake is formed on the trailing side. Although   
the wake should exert a drag force on the compact object, 
the drag force weakens rapidly (approximately following $v_*^{-2}$) as $v_*$ increases \citep{1999ApJ...513..252O}. For high-velocity passages, the slow-down of the compact object by the drag force is not significant. 

% part from the drag force, 
%For the wake to contract into a contrail, the encounter should inject enough momentum to trigger collapse.

Within $r_{\rm wake}$, the passage injects a momentum that is comparable to the velocity dispersion of the medium. As a result, we expect the gas to contract which leads to a significant amount of density enhancement.  This density enhancement has a collimated, contrail-like morphology, and the contrail should be spatially aligned with the orbit of the object measured in the rest frame of the cloud, thus we name it as ``molecular contrail''. 

% The contrail is the very result of the contraction of the wake. Without detailed knowledge on the contraction process, we make the working hypothesis that the contractions do not produce order-of-magnitude enhancements in the density of the molecular gas, i.e. the contrail width $d$ is comparable to the estimated radius of the wake, $d \approx r_{\rm wake}$.

\begin{figure*}
  \includegraphics[width = 0.95 \textwidth]{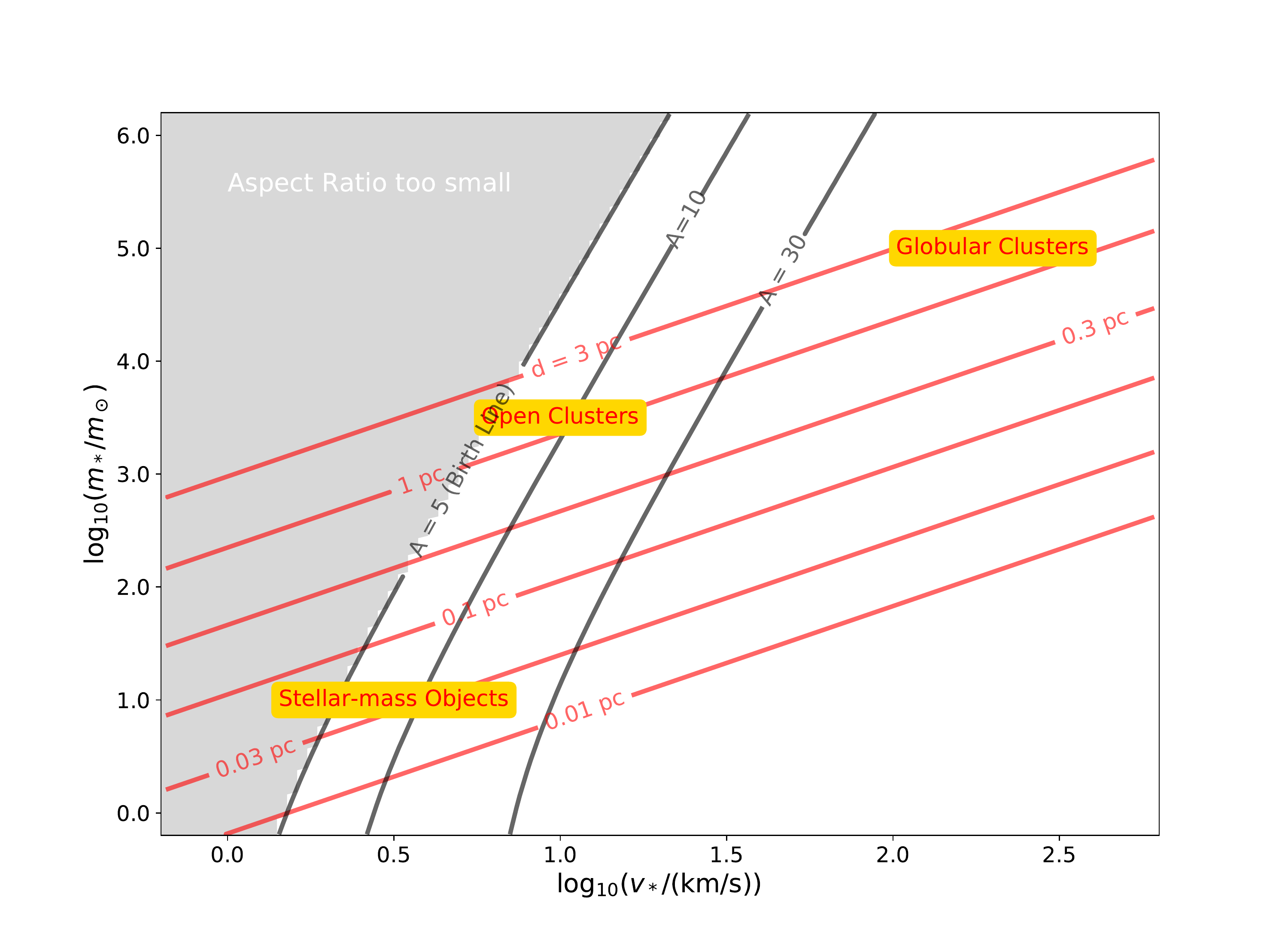}
  \caption{Contrail formation in the Milky Way. The $x$-axis
    is the logarithm of the velocity of the compact object
    measured with respect to the molecular cloud or clump, and the $y$-axis
    is logarithm
    of the mass of the object. Different initial conditions lead to different contrail widths and aspect ratios, which are indicated by the red and black lines, respectively. The gray shaded region indicate the parameter range where the contrails have aspect ratios that are too small to be identified. Our candidate contrail-producing objects include stars, star clusters as well as globular clusters.  Their locations are indicated in the plot. See Sec. \ref{sec:occurence} for details. \label{contrail:production}}
\end{figure*}

We are only interested in fast passages where the encounter velocity $v_*$ is larger than the velocity dispersion of the medium, since this is a necessary condition for the contrail to stay aligned spatially. We note that after  formation, the contrail is subject to destruction by the velocity dispersion of the medium. Considering that the contrail can be destroyed in a crossing time by the random motion of the molecular cloud, the survival time of a contrail of width $d$ is
\begin{equation}
  t_{\rm survival} \approx d / \sigma_{\rm medium}\,.
\end{equation}
Thus, $v_* >>  \sigma_{\rm medium}$ is equivalent to the aspect ratio of the contrail, which can be estimated by
% , $t_{\rm inject} \ll r_{\rm wake} / \sigma_{\rm medium}$, i.e. $v_* \gg \sigma_{\rm medium}$.
%The aspect ratio of a contrail can be estimated by
%is physically determined by
\begin{equation}\label{eq:a}
  A = v_*\, t_{\rm survival} / d % \approx v_* / \sigma_{\rm wake}
  = v_* /   \sigma_{\rm medium} \;,
\end{equation}
is much greater than unity.

From these calculations we can conclude that, in general, compact objects with higher velocities lead to more collimated contrails, and those with higher masses lead to contrails of larger widths. Both the width and the aspect ratio of the contrails are inversely proportional to the velocity dispersion of the molecular clouds.

\subsection{Formation condition in the Milky Way}\label{sec:larson}
It has been established observationally that in the Milky Way, the turbulent motion of the molecular gas is described by the Larson's relation \citep{1981MNRAS.194..809L}, where the velocity dispersion $\sigma_{\rm turb}$ of the medium is related to the scale $l$ by $\sigma_{\rm turb} \sim \ell^\beta$, $1/3 <= \beta<=1/2$ \citep[e.g.][]{2009ApJ...699.1092H}. We adopt $\beta=1/3$ which is consistent with the Kolmogorov turbulence spectrum, and write the total velocity dispersion of the molecular gas as
\begin{equation}\label{eq:larson}
  \frac{\sigma_{\rm medium}}{1\;\rm km/s} = \bigg{(} \Big{(}\frac{\ell}{\rm 1\;\rm pc}\Big{)}^{2/3} + \Big{(}\frac{c_{\rm s}}{1\;\rm km/s}\Big{)}^2 \bigg{)}^{1/2}\;.
\end{equation}
%where the first term describes the dependence of velocity dispersion on the scale as found in  \citep{1981MNRAS.194..809L},
The first term in the parentheses describes the turbulence contribution as found in \citet{1981MNRAS.194..809L}, and
the second term describes
 the contribution from the sound speed
  $c_{\rm s}$
  of the medium.
For contrail formation and disruption, the relevant scale of the turbulence is $\ell \approx r_{\rm wake} \approx d$. As the typical temperature of molecular cloud in the Milky Way is $T \sim 10 \;\rm Kelvin$ corresponding to a sound speed of $c_{\rm s} \approx 0.2 \; \rm km/s$. The sound speed of the isothermal molecular cloud is computed as $c_{\rm s}=\sqrt{k_{\rm B} T / \mu m_{\rm H}}$, where $k_{\rm B}$ is the Boltzmann  constant, $\mu = 2.7 $ is the mean molecular weight, and $m_{\rm H}$ is the hydrogen atomic mass.
%the turbulence contribution dominates $\sigma_{\rm medium}$ for contrails with $d \gg 0.008 (T/10\;\rm K)^{3/2}$\;pc. Therefore
The sound speed contribution can be safely neglected for $d\gtrsim 1$pc contrails, and is only relevant for thin contrails such as the $d \lesssim 0.01$ pc  (where $\sigma_{\rm turb} < c_{\rm s}$)  contrails.  

Combining this expression of $\sigma_{\rm medium}$ with Eqs. \ref{eq:d} and \ref{eq:a}, we extract a diagram of contrail formation (Fig. \ref{contrail:production})
describing the dependencies of the contrail width and aspect ratio on $m_*$ and $v_*$.
In the large $d$ limit where the turbulence contribution dominates $\sigma_{\rm medium}$, we have
\begin{equation}
  \frac{d}{1\;\rm pc} \approx
  0.017 \, \Big{(}\frac{m_*}{m_{\odot}} \Big{)}^{3/4} \Big{(}\frac{v_*}{\rm 1\, km/s} \Big{)}^{-3/4}\;,
\end{equation}
and
\begin{equation}
  A \approx 3.8 \; \Big{(} \frac{m_*}{m_{\odot}} \Big{)}^{-1/4 }\, \Big{(} \frac{v_*}{\rm 1\,km/s}\Big{)}^{5/4}\;.
\end{equation}

As examples, we consider the following types of contrails:  the $d\gtrsim 0.01\;\rm pc$ contrails which can be observed with interferometers, as well as the  $d\gtrsim 1\;\rm pc$ contrails which should be observable by single-dish telescopes. In addition to this,
if we demand that the minimum aspect ratio of a contrail to be $A\sim$ 5 (which we consider as the ``birth line'' for
contrails, see Fig. \ref{contrail:production}), we can derive the minimum masses and impact velocities for contrails
of different widths: For example, to form a $d \gtrsim 0.01\;\rm pc$ contrail, %like the W43 contrail and the G33.94 contrail,
one requires at least a stellar-mass object, and for
a $d \gtrsim 1\;\rm$ pc contrail,
%like the W51 contrail,
at least a 1000 $m_{\odot}$ object. % such an open cluster or an intermediate-mass black hole   \citep[][and references therein]{2004IJMPD..13....1M}.
These minimum masses at the birth line are associated with typical minimum impact velocities of a few km/s.
Contrails of the same widths but potentially higher aspect ratios
% \footnote{Observed aspect ratios of the molecular contrails can be affected by the structure of molecular clouds and cease to reach the estimated value.}
can be formed by objects with higher masses and velocities but the same $m_*/v_*$ ratio. For instance, for compact objects with velocities on the order of the virial velocity of the Galactic halo, $v_* \gtrsim 10^2 \; \rm km/s$, forming $d \gtrsim 0.01\;\rm pc$ contrails requires $m_* \sim 10^2 m_{\odot}$ and forming $d \gtrsim 1\;\rm pc$ contrails requires $m_* \sim 10^5 m_{\odot}$.

\subsection{Formation routes and theoretical formation rates}\label{sec:occurence}

% For convenience, we consider two set of situations where contrails are produced. Stellar-mass objects are responsible for contrails of $d \approx  0.01 \;\rm pc$ whereas star clusters are responsible for contrails of $d \approx 1 \;\rm pc$.
%
% \xun{\sout{\xun{\sout{Observationally, there are two types of contrails} The contrails we identified in the observations fall into two groups}: One is the
% $d \gtrsim 0.01\;\rm pc$  contrails identified from interferometer observations of the disk, such as the W43 contrails. These observations tend to target at pc-sized regions of high densities, and these contrails are typically found in regions where clustered star formation are taking place. \xun{\sout{Ideally, $d \gtrsim 0.01\;\rm pc$}In principle, these thin} contrails can also be seen in regions of lower densities. However, because of a lack of observations, we can not study them in detail at this stage \footnote{These contrails are also ``less interesting'' since in contrast to the contrails identified from dense regions, they reside in  low-density regions and are expected to have low criticality (see Sec. \ref{sec:criticality}). As a result we do not expect them to  collapse further under self-gravity.}.
% Another type of contrails are the $d \gtrsim 1\;\rm pc$ contrails identified from mapping observations of the Galactic disk. They can be produced by passages of star clusters (including open clusters and globular clusters) through molecular gas.}}
{bf 
The dense objects which can cause contrails include stars, star clusters as well as black holes of different masses. }
To estimate the occurrence rates of different contrails, we consider an ensemble of compact objects
of a number density of $n_{*}$ and a velocity dispersion of $\sigma_{\rm v *}$, and consider their encounter with  molecular clouds/clumps of size $\ell_{\rm c}$. The mean free path of the encounter is
\begin{equation}
  \lambda_{\rm mfp} = 1/(n_* \ell_{\rm c}^2)\;,
\end{equation}
and the typical encounter time is
\begin{equation}
  t_{\rm encounter} = \lambda_{\rm mpf} / \sigma_{\rm v *} = 1/(n_* \ell_{\rm c}^2 \sigma_{\rm v *})\; .
\end{equation}
%To estimate
Then, the occurrence rate can be estimated as the ratio of contrail survival time and the encounter time,
\begin{equation}
  t_{\rm survival}/t_{\rm encounter} = n_*\, d\, \ell_{\rm c}^2\, (\sigma_{v*} / \sigma_{\rm medium})\;.
\end{equation}
We use this equation to estimate the occurrence rates of contrails formed through different channels.

\subsubsection{Stellar passages}
First we consider the formation of contrails of $\gtrsim 0.01 \;\rm pc$ by
passages stellar-mass objects through molecular clumps. The clumps are pc-sized
regions of high densities 
\citep[e.g. $n_{\rm H_2}\gtrsim 10^4 \;\rm cm^{-3}$ for a typical, $10^3 M_{\odot}$ clump, ][]{2018MNRAS.473.1059U} and they have  steep density profiles $\rho_{\rm clump} \approx r^{-2}$ \citep[][and references therein]{2018MNRAS.477.4951L}.
{  To produce the $d\gtrsim 0.01 \;\rm pc$ contrails, we require a minimum mass of 10 $M_{\odot}$ and a velocity of $30\;\rm km/s$. 
Out candidates are mostly located in the innner part of the Galactic disk. Using results from \citet{2016ApJ...816...42M}, we estimate a stellar mass surface density of $300\;M_{\odot}\;\rm pc^{-2}$ at $r_{\rm gal} = 4 \;\rm kpc$ where $r_{\rm gal}$ is the galactocentric distance. From this, we estimate a volume mass density of $2 M_{\odot}\;\rm pc^{-3}$. Assuming a standard IMF \citep{1955ApJ...121..161S}, we estimate a number density of 0.04 pc$^{-3}$ and a mean free path of 25 pc for stars with $M \gtrsim 10 M_{\odot}$. This implies a encounter time of $t_{\rm encounter}=30 {\rm pc }/ 30 {\;\rm km/s} = 0.8 \;\rm Myr$.  The survival time is $t_{\rm
      survival} \approx 0.01 \;\rm pc / (0.3 \;\rm km/s)  \approx 0.03 \; \rm Myr$.
      We thus expect
to be able to observe contrails in around  $t_{\rm survival} / t_{\rm encounter}=$4    \% of the clumps. 

{  Note that apart from gravity, massive stars can influence the ambient ISM
thorough radiation pressure and photoionization. At high densities, radiation pressure dominates the momentum feedback \citep{2009ApJ...703.1352K}. For massive stars
, the radiation pressure approach
the Eddington limit when $m_* \gtrsim 40 M_{\odot}$ \citep{2015A&A...580A..20S} such that the momentum feedback on the ambient medium becomes non-negligible. However, as the number density of stars decreases very rapidly as the stellar mass increases, the radiation pressure is relevant only to a small fraction of the cases and that we are concerned with. 
}

% {\color{red}{BH? ``The estimated number of black holes in the Galaxy from simple stellar evolution considerations is about 100 million, a large fraction of which are expected to besingle. Yet, not a single isolated black hole has been detected to date -- all the few dozen black hole mass determinations so far have been in binaries. In addition, there is a nagging inconsistency between the measured masses of black holes in our Galaxy, masses expected from theoretical calculations, and the LIGO measurements. Mass determinations of a few isolated, stellar-mass black holes will provide important clues in our understanding of black holes. Astrometric microlensing is the only available technique capable of detecting isolated black holes and measuring their masses.''
% HST program: https://ui.adsabs.harvard.edu/abs/2012AAS...22030703S/abstract
%

{  

\subsubsection{Passages of stellar-mass black holes}
The detections of gravitational waves by the Laser Interferometer Gravitational-Wave Observatory (LIGO)
 has revealed the existence of a population of binaries  where the mass of the black hole can reach a few tens of solar masses \citep{2016PhRvL.116x1103A,2017PhRvL.118v1101A}. These stellar-mass black holes are massive enough to cause contrails. The total number and the mass distribution of the stellar-mass black holes in the Milky Way remain uncertain. 
  According to recent models \citep[e.g.][]{2017MNRAS.468.4000C,2018MNRAS.473.1186E}, the Milky Way may contain up to $10^8$ black holes whose masses are larger than $10 M_{\odot}$. These black holes should be able to trigger the formation of a significant number of $d \gtrsim 0.01\;\rm pc$  contrails. 

We estimate the occurrence rate of contrails caused by black hole passages. 
The spatial distribution of these black holes is not known yet. We assume that they follow a distribution which resembles that of the stellar disk. Using results from \citet[][]{2018MNRAS.473.1186E}, and assuming a volume of $10{\ \rm kpc}\times 10{\ \rm kpc}\times 0.3 {\ \rm kpc} $ and a velocity dispersion of $\rm 30\; km/s $ for these objects,  we find  a mean free path of $l_c\approx 300\;\rm pc$ and a collision time of 10.0 $\rm Myr$, and these passages can produce contrails of $d \gtrsim 0.01\;\rm pc$.  If the contrail survives for a crossing time, which  is around 0.03 Myr, we estimate that that 0.3       \% of the clumps should contain contrails of this type. 
}}

\subsubsection{Passages of star clusters }
Next, we consider the production of contrails of $\gtrsim 1\;\rm pc$ by passages of
star clusters through molecular clouds in the Galactic disk. {  Stellar associations can shape the structure of the ambient ISM through radiation \citep[e.g.][]{2018A&A...619A.120K}.} Passages of star clusters through clouds are regular events, and such passages can contribute to the destruction of star clusters \citep{2006MNRAS.371..793G}.  {These star clusters have a typical size of 1--2 pc \citep{2009ApJ...705..468H}, which is comparable to the width of the contrails that they will be creating}.
From some recent observations \citep[e.g.][]{2019AJ....157...12B}, the number density of star clusters is estimated to be $n_{*} \approx 2 \times 10^4  \;\rm kpc^{-3}$.
With $\ell_{\rm c} = 10 \; \rm pc$ and $\sigma_{\rm v *} = 30\;\rm km/s$, the  mean free path is estimated to be 0.5 $\rm kpc$ and the encounter time is around  $15\;\rm Myr$. The survival time for such contrails is around $t_{\rm survival} = 3 \;\rm pc / (1\;\rm km/s)  \approx 3 \; \rm Myr$
, the chance of seeing such contrails  in a molecular cloud is around 20\%.
% \xun{a table?}

\subsubsection{Passage of globular clusters }
We consider the production of the contrails by passages of globular
clusters through the Galactic disk. Globular clusters have typical
masses of 10$^5$ - 10$^6$ $M_{\odot}$ and typical velocities measured with respect to the molecular clouds of $10^2$ km/s. They are thus capable of producing  $\sim 1$ pc wide contrails.
{Globular clusters are far from point masses,   but they have concentrated/cored density profiles. Their density profiles can be described approximately as
$\rho \sim \rho_0 / (1 + (r/r_{\rm c})^2) $ \citep{1962AJ.....67..471K}  { and their half-mass radius range from sub-pc to a few parsec \citep[e.g. Table 4 of][]{2016A&A...587A..53K}.} When such globular
clusters pass through a cloud to produce the 1 pc-wide contrails, only the stars at the very center of the clusters are effective in injecting momentum.}
%The globular clusters orbit around the Milky Way and pass through the disk at rate that is inversely proportional to the dynamical time $t_{\rm dyn}$.
%Measured with respect to the molecular clouds whose motions are mostly Keplerian, their typical impact velocities are comparable to  the
%Keplarian velocity e.g. 100 km/s.
Around $N_{\rm GC} \approx 10^2$ globular cluster candidates have
been identified around the Milky Way halo \citep{2019MNRAS.487.3140P}.
They pass through our Galaxy every $t_{\rm orbit} \approx 10^8$ yr.
Due to this relatively small number, rather than estimating the contrail occurrence rate for each molecular cloud, we estimate the  total number of contrails associated with globular clusters using % in the molecular disk,
$N_{\rm contrail}^{\rm GC}  \approx  t_{\rm survival}/t_{\rm orbit}  \times N_{\rm encounter} $, where
%$t_{\rm cross}$ is the disk-crossing time of roughly $100\;{\rm pc} / (100\;\rm{km/s})$,
$t_{\rm survival} \approx 1\;{\rm pc} / (1\;\rm{km/s})$,
%rate at which globular clusters pass through the disk,
and $N_{\rm encounter}$ is the number of clouds a globular cluster encounters at a single passage through the Galactic disk. Assuming $N_{\rm encounter}=10$, we expect to see a few of these contrails in the Milky Way.
%$t_{\rm cross}$ is the disk-crossing time of roughly $100\;{\rm pc} / (100\;\rm{km/s})$,
Due to the high relative speed, the globular cluster encounters should be responsible for the production of some of the  longest and most collimated contrails.

%a few tens of contrails produced by globular clusters in the whole galactic disk. {\color{red}{$t_{\rm survival} / t_{\rm cross}$ rather than $t_{\rm dyn}$}
%Because of the high impact velocity, these contrails should be highly collimated.

 \begin{figure*}
  \includegraphics[width = 1 \textwidth]{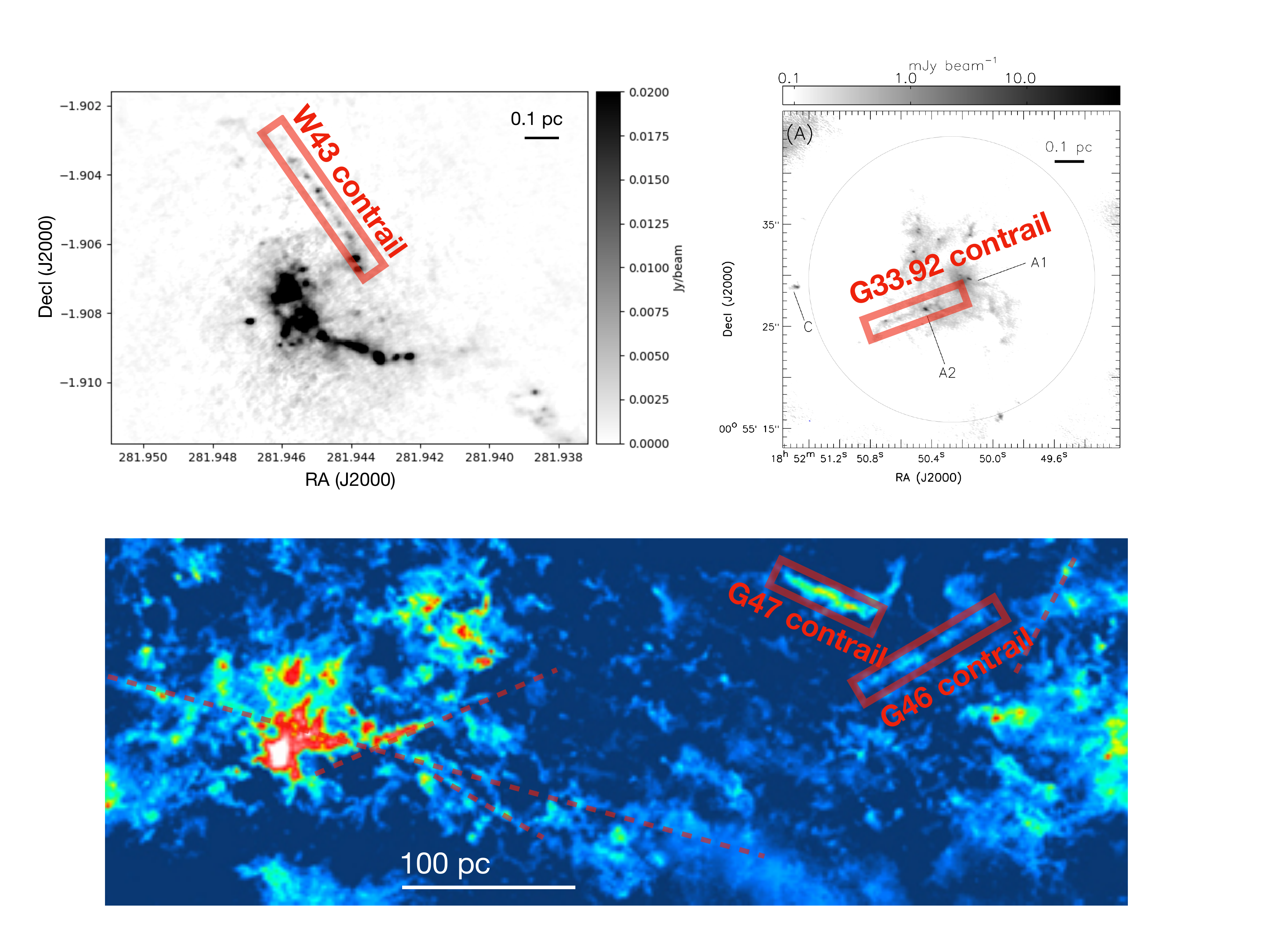}
  \caption{\label{fig:1}
  Maps of molecular contrails of various { lengths and widths }in different {Galactic} regions. \textbf{Upper left}: ALMA 1.3mm continuum observation of the W43 massive star-forming
  region from \citet{2018NatAs...2..478M}, where dense gas are traced using emission from dust. {  Reproduced using data shared by the authors.}
  \textbf{Upper right}: ALMA 1.3mm continuum observation of the high-mass star-forming clump G33.92$+$0.11 from \citet{2019ApJ...871..185L}. {  Reproduced with permissions from the authors and the AAS.}
  \textbf{Bottom panel}: velocity-integrated $^{13}$CO(1-0) map of a section of the Milky Way (46$^{\circ}$<$l$<50$^{\circ}$, $-1^{\circ}<b<0.5^{\circ}$) integrated {from $v_{\rm LSR}$=33 to 70 km/s}. The data are obtained from the Galactic Ring Survey (GRS) \citep{2006ApJS..163..145J}.
  The gas in the plot sits at a distance of $\sim$ 8 kpc. In all panels, major contrail candidates are indicated using red boxes marked with their names.
  In the bottom panel, a few other contrail candidates are indicated using red dashed lines. {  Reproduced with permissions from the authors and the AAS.}}
\end{figure*}

\subsection{Criticality and Collapse}
\label{sec:criticality}
{ The criticality of a contrail can be estimated roughly from the ratio between gravitational and turbulence energy densities,  } \citep[][]{1953ApJ...118..116C,1963AcA....13...30S,1964ApJ...140.1056O}
\begin{equation}\label{eq:crit:def}
  \delta_{\rm contrail} = \frac{G \;M/L}{2 \;\sigma_{\rm medium}^2 } \;,
\end{equation}
where $M/L$ is the line mass and $\sigma_{\rm medium}$ is the estimated velocity dispersion of the contrail measured at the radial direction \footnote{It contains possible contributions from compression, shear, etc.}.
Contrails with $\delta_{\rm contrail} \gg 1$ are prone to fragmentation. Note that  Eq. \ref{eq:crit:def} also suffers from uncertainties arising from  i.e. the equation of state of the gas and the radial density profile, and {  the measured values of $\delta_{\rm contrail}$ are also affected by the inclination angle.}

\section{Observational signatures}
\subsection{Contrail candidates}
\label{sec:obs}
\begin{figure*}
  \includegraphics[width = 0.95 \textwidth]{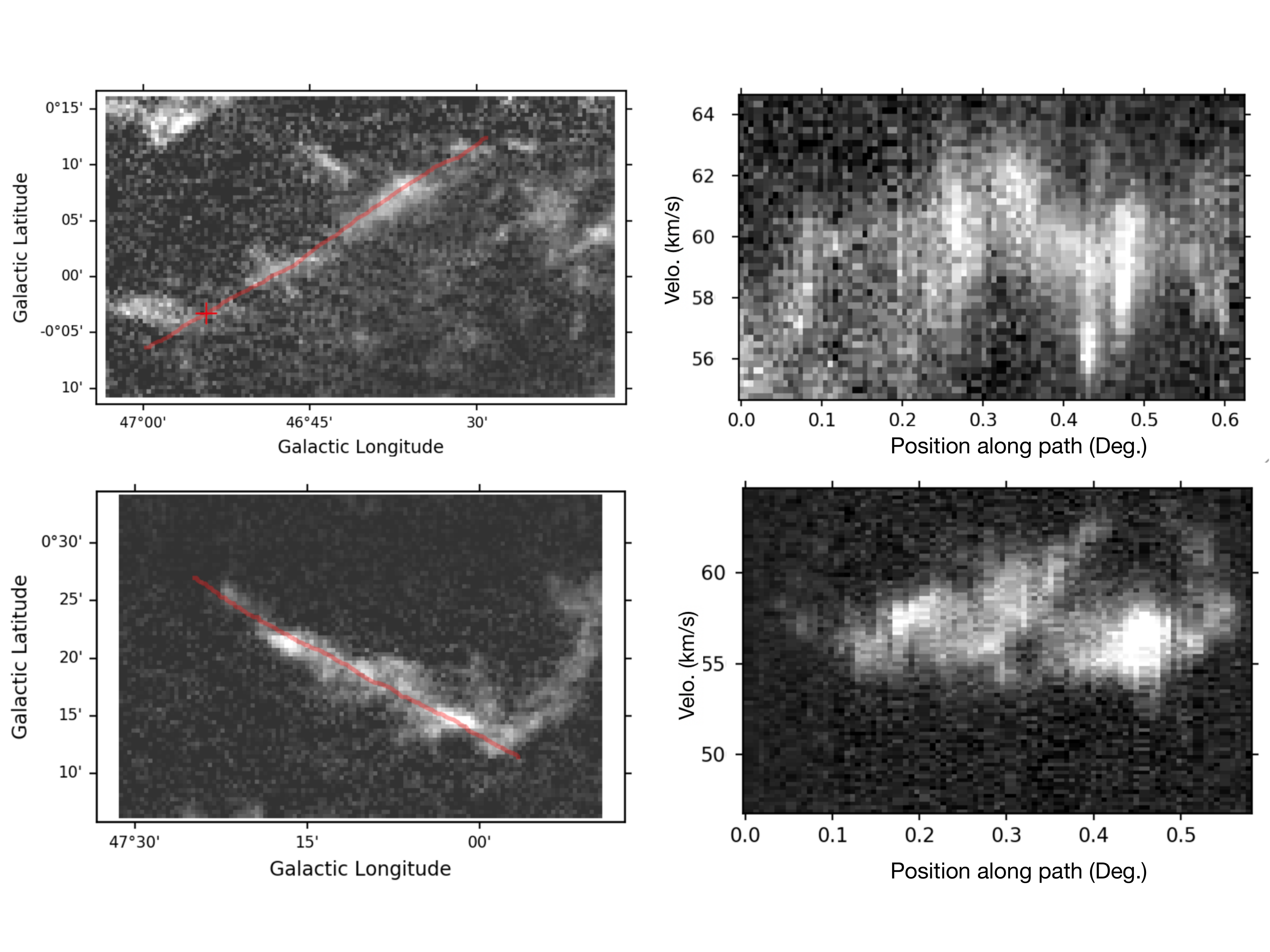}\\
  \caption{ Upper panels: image and velocity structure of G46 contrail. The upper left panel is a map of $^{13}$CO(1-0) map obtained at $v_{\rm LSR} =\rm 60 km/s$. The upper right panel is a position-velocity diagram of the same contrail, where it is taken from the slice denoted by the red line in the left panel. Lower panels: map and velocity structure of the G47 contrail. The map is taken at  taken at $v_{\rm LSR} =\rm 56 km/s$. \label{fig:velo}   }
\end{figure*}

In Fig. \ref{fig:1}, we present a few contrails identified from
published observational data on the surface density distribution of molecular gas in different regions.
The contrails in the upper panels (W43 contrail, G33.92 contrail) are identified from dust continuum observations towards high-mass star-forming regions \citep{2018NatAs...2..478M,2019ApJ...871..185L} carried out using the ALMA telescope \citep{2009IEEEP..97.1463W}.  {In these interferometer observations, because of the unavoidable incompleteness of the UV coverages, only structures with significant, localised surface density variations are recovered.} The lower panel is constructed using the $^{13}$CO(1-0) data from the Galactic Ring Survey (GRS, \citet{2006ApJS..163..145J}) carried out by the  Five College Radio Astronomy Observatory
(FCRAO) 14m telescope, in which we have identified the G46 contrail, the G47 contrail, together with a few additional contrail candidates. { The velocity structure of some contrails are presented in Fig. \ref{fig:velo}.}
The estimated parameters for the four contrails %candidates
with names are listed in Table \ref{table:1}. Details concerning how we derive
these values can be found in Appendix \ref{sec:estimation}. Since the contrails
do not have well-defined boundaries, these values are only rough estimates.
Limited by the line of sight confusion, we are not able to measure the velocity
dispersion of the contrails directly. The velocity dispersions of the contrail
in the radial direction are inferred assuming that the medium is dominated by a
turbulent motion described by Eq. \ref{eq:larson}.

These contrails share a striking similarity in that they consist of %a set of
spatially-aligned density enhancements, which leads to 
high aspect ratios ($\sim 10$). They exist, however, on a wide range of spatial
scales, with widths ranging from $\sim 0.01\;\rm pc$ to a few pc and lengths
ranging from 0.1 pc to $10^2$ pc.    Using data from the GRS survey, we find
that the G46 and G47 contrails seem to exhibit coherent yet turbulent velocity
structures (Fig. \ref{fig:velo}).  %a few tens of parsec. %and a common
% character shared by these contrails is that they have high aspect ratios  ($\sim
% 10$).
The contrails also exhibit different degrees of fragmentation. The more
``fragmented'' W43 contrail is basically a linear chain of dense cores.

Considering the population of compact objects in the Milky Way and their typical velocities, $d \gtrsim 0.01\;\rm pc$ contrails like the W43 contrail and the G33.92 contrail are most likely formed by massive stars, binaries, and  stellar-mass black holes, and $d \gtrsim 1\;\rm pc$ contrails like the W43 contrail and the G33.92 contrail are most likely formed by star clusters.

Since our sample is still limited, the occurrence rate of the contrails is not well constrained.  Nevertheless, the appearance of $d\approx 1\;\rm pc$
contrails do appear to be a regular event.  We have identified five contrails and
plausible contrail candidates from the data shown in the bottom panel of Fig. \ref{fig:1},
and all of them are longer than 50 pc. As a comparison, according to
\citet{2009ApJS..182..131R}, the region contains a total of 40 molecular clouds.
Most of these clouds  are small in size, and only half of the clouds have sizes
larger than 50 pc such that they can potentially host these contrails. Based on
this, at least 20 \% of the large clouds in the region host
contrails. Better
constraints on the contrail occurrence rate will be derived in future works.

\begin{table*}
  \begin{tabular}{lrrrrrrr}
    \hline
    Name            & Length $L$ & Width $d$ & Aspect ratio $A$ & Mean surface density    & M/L                    & $\sigma_{\rm medium}$ & $\delta_{\rm crit}$ \\
    \hline
    W43 contrail    & 0.5 pc     & 0.02 pc   & { 24}            & 4 $\rm g\, cm^{-2}$     & 400 $M_\odot/\rm pc$   & 0.37 km/s             & 5                   \\
    G33.92 contrail & 0.2 pc     & 0.01 pc   & 20               & 0.6  $\rm g \,cm^{-2}$  & 27  $M_\odot/\rm pc$   & 0.36 km/s             & 0.5                 \\
    % W51 contrail    & 25 pc    & 3 pc      & 8                & 0.06 $\rm g\, cm^{-2}$  & 800 & 1.7 km/s              & 0.9                 \\
    G46 contrail    & 80  pc     & 3 pc      & 10               & 0.02  $\rm g\, cm^{-2}$ & 300  $M_\odot/\rm pc$  & 1.5 km/s              & 0.3                 \\
    G47 contrail    & 60  pc     & 3 pc      & 6                & 0.03  $\rm g\, cm^{-2}$ & 450   $M_\odot/\rm pc$ & 1.5 km/s              & 0.4                 \\

    \hline
  \end{tabular}
  \centering
  \caption{List of contrail candidates and their physical properties. See Appendix \ref{sec:estimation} for details. \label{table:1}
  }
\end{table*}

\subsection{Velocity structure}
Using publicly-available data from the GRS survey \citep{2006ApJS..163..145J}, we construct the position-velocity diagram of two $d\approx1\;\rm pc$ contrails (Fig. \ref{fig:velo}). The contrails appear to be coherent in the position-velocity space, confirming that our contrails are coherent objects. One can also see that these contrails still contain significant amounts ($\approx 3\;\rm km/s$) of line-of-sight turbulent motion.

This is barely surprising since the observed velocity distributions
contain contributions from both the turbulent cloud (roughly a few tens of pc in size)
and contributions from       the contrails ($\approx 1$ pc in size). From Eq. \ref{eq:larson},
we estimate a velocity dispersion of a few km/s for the background cloud and a velocity dispersion of around $1\;\rm km/s$ for the contrail. Since the background cloud contains a velocity dispersion  that is much larger than that of the contrail, the velocity structure seen in Fig.
\ref{fig:velo} is dominated by the longitudinal turbulent motion. This explains the fact that  the position-velocity diagrams of our contrails do not appear to be
different from those of ordinary molecular clouds.

\subsection{Criticality}

\begin{figure}
  \includegraphics[width = 0.55 \textwidth]{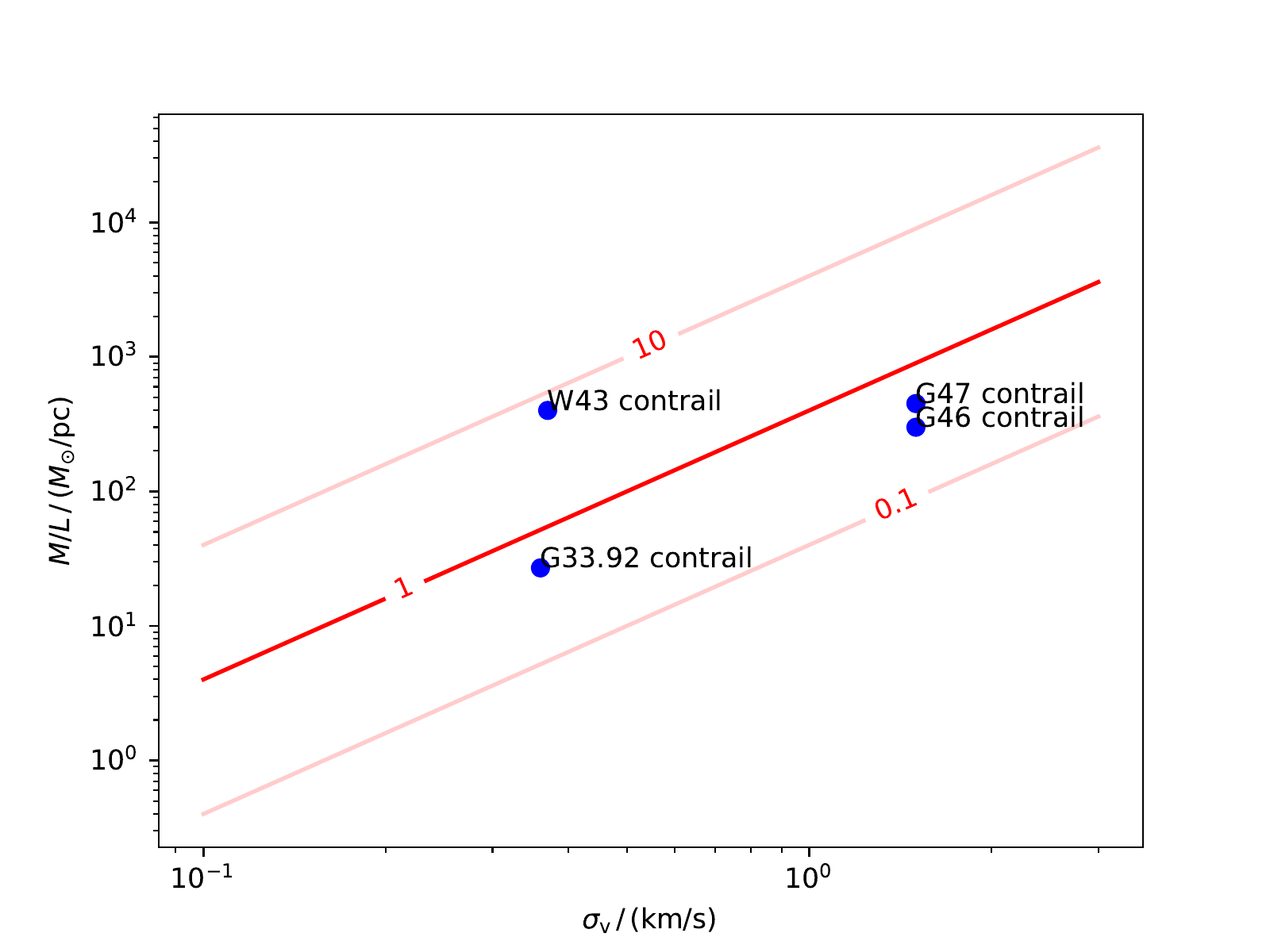}
  \caption{\label{contrail_critality} Contrail line mass (M/L) plotted against their velocity dispersions. Lines that correspond to different criticality parameters $\delta_{\rm contrail}$ are indicated.  }
\end{figure}

In Fig. \ref{contrail_critality}
we present a criticality diagram, where most of our contrails stay close or above the critical line defined as $\delta_{\rm contrail} \approx 1$, meaning that the gravitational energy density is comparable or larger than the kinetic energy density of the turbulent motion. Contrails with even smaller criticality parameters should exist in principle, but  are probably more difficult to identify. The only contrail which
has a very high critical parameter is the W43 contrail, where  $\delta_{\rm
    contrail} \approx 10$. This corresponds well to the fact that it is the one
with the highest degree of fragmentation.

\section{Discussions and Conclusions}
% We report the ubiquitous existence of straight, beam-like concentration of gas in the molecular ISM, and call them ``molecular contrails". We found that molecular contrails in general have large aspect ratios. They exists in different sizes with diverse morphology. In certain instances, contrails are found to be associated with high concentration of dense clumps and cores.
%
%
% We find that molecular contrails can be easily produced the passages compact objects through clouds, where gravity injects momentum into the medium which drives the compressions. Presumably, the narrower (d $\approx$ 0.01 pc) contrails are presumably produced by passages of stars through the clouds, where as the wider (d $\approx$ 1 pc)  contrails are produced passages of star clusters and intermediate-mass black holes. Certain contrails are super-critical which means they should collapse to form a new generation of stars.
%

% properties of those second generation of star clusters might be
%statistically different from that of the first generation. The the passage of massive
%stars through clouds might induce the formation of a stellar population with a
%different initial mass distribution. The mechanism we proposed can have far-reaching
%consequence which deserves further investigation

The Milky Way is a complex system made of gas, stars and dark matter. We investigate the effect of compact objects such as stars {  black holes} and star cluster on the molecular gas when they pass through it, and find that fast passages -- passages where the relative velocity is much larger than the velocity dispersion of the medium, leads to the formation of spatially aligned concentration of dense gas called the molecular contrails.  We derive the necessarily conditions for the formation of contrails of different widths.
 According to our analysis, contrails of $d\gtrsim$  0.01 pc
are caused by passages of stellar-mass objects,  and contrails of
$d\gtrsim$ 1 pc are caused by passages of star clusters. We predicted that the formation of these contrails, especially the $d\gtrsim$ 1 pc ones, should be a regular event.

From observational data,  we identify a set of spatially-aligned concentrations of dense molecular gas of difference widths and sizes, and propose that they are the molecular contrails predicted by our analysis. The contrails are also sensitive to the existence of other massive objects
and can be used to trace them.

Our findings have revealed a seldomly-noticed channel of triggered collapse of the molecular gas. We note that \citet{1996ApJ...459..555W} have previously
proposed that the passage of globular cluster can trigger star
formation. We further point out that that this should occur
in the form of molecular contrails.
Since the formation of the dense molecular gas is considered as a key step towards star formation, our findings have effectively revealed a new way to trigger star formation in galaxies.  Molecular contrails may contribute significantly to the total star formation rate under certain conditions, i.e. when self-gravity is not sufficient to cause the gas to collapse.  This applies to molecular clouds in gas-poor elliptical galaxies \citep{2018ApJ...858...17T} and  the Galactic center \citep{2020ApJ...897...89L}.
Knowledge about the formation and evolution of contrails should thus be an inseparable part of our understanding of the star formation process as a whole, which we expect to gain through future efforts.\\

\section*{Acknowledgements}
We thanks our referee a very careful reading of our paper and the valuable comments. {  Guang-Xing Li sknowledges supports by NSFC grant W820301904. We thank Frederique Motte and authors of \citet{2018NatAs...2..478M} for sharing their data. We thank Dr.  Hauyu Baobab Liu and the AAS for allowing us to reuse Fig. 1 of \citet{2019ApJ...871..185L}. We thank authors of \citet{2006ApJS..163..145J} and the AAS for allowing the reusage one of their Figures.}
%%%%%%%%%%%%%%%%%%%%%%%%%%%%%%%%%%%%%%%%%%%%%%%%%%
\section*{Data Availability}
The data used to produce the bottom panel of Fig. 2 and Fig. 3 are available at \url{https://www.bu.edu/galacticring/new_data.html}.

\bibliographystyle{mnras}
\bibliography{paper}
%%%%%%%%%%%%%%%%%%%%%%%%%%%%%%%%%%%%%%%%%%%%%%%%%%

\appendix

\section{Estimation of  other contrail parameters} \label{sec:estimation}
The estimated parameters for the four contrails %candidates
with names are listed
in Table \ref{table:1}. The length $\ell$, widths $d$ and mean surface densities
are measured directly from the data. Note that (a) the contrails exhibit
density variations along its ridges, such that the definition of contrail
length can be vague.
(b) the contrail widths are defined by the FWHM of the intensity distribution
on the observation maps. Limited by resolutions of available observations, an accurate determination
of the density profiles of these contrails is not possible,
and the contrail widths could be over-estimated by a small margin. The aspect ratio is taken as the ratio between filament length $\ell$ and width $d$, {\ and in some cases, their aspect ratios are reevaluated using method described in
    in  Appendix \ref{sec:aspect ratio}.}
(c) while
measuring the contrail surface densities from the ALMA maps, we have used the
conversion factors adopted by the authors. To compute the surface density from
the GRS \citep{2006ApJS..163..145J} observations, we have used the conversion factor
from  \citet{2001ApJ...551..747S}. From these directly measured quantities, we
have further derived several parameters important in the proposed physical picture of
contrail formation, including the line mass
$M/L$, the velocity dispersion $\sigma_{\rm medium}$ of the molecular gas measured on a
scale that is comparable to the contrail width (see Section
\ref{sec:larson}), and a criticality parameter
$\delta_{\rm contrail} = G (M/L) /  2 \sigma^2_{\rm medium}$ (see Section
\ref{sec:criticality}). The line mass is estimated
using $M/L \approx \Sigma_{\rm gas} \times d$ where $\Sigma_{\rm gas}$ is the
estimated surface density. % and $d$ is the estimated width.
The velocity
dispersion $\sigma_{\rm medium}$ of the ambient molecular gas on scale $d$ is
dominated by turbulence dispersion for large
$d\gtrsim 1\;\rm pc$ (W51 contrail and G46 contrail), and the thermal velocity dispersion is important only for small $d \gtrsim 0.01 \;\rm pc$ contrails (W43 contrail and G33.92 contrail). Thus, we estimate the
$\sigma_{\rm medium}$ for the W51 contrail
and the G46 contrail using the empirical Larson relation (see Section
\ref{sec:larson} for details), and that for the W43 contrail and the G33.92 contrail using additional information of molecular gas temperatures
estimated in the corresponding papers (23K for the  W43 contrail and 30K for the G33.92 contrail).

\section{Contrail separation and measurement of aspect ratio}\label{sec:aspect ratio}
\begin{figure*}
  \includegraphics[width = 1 \textwidth]{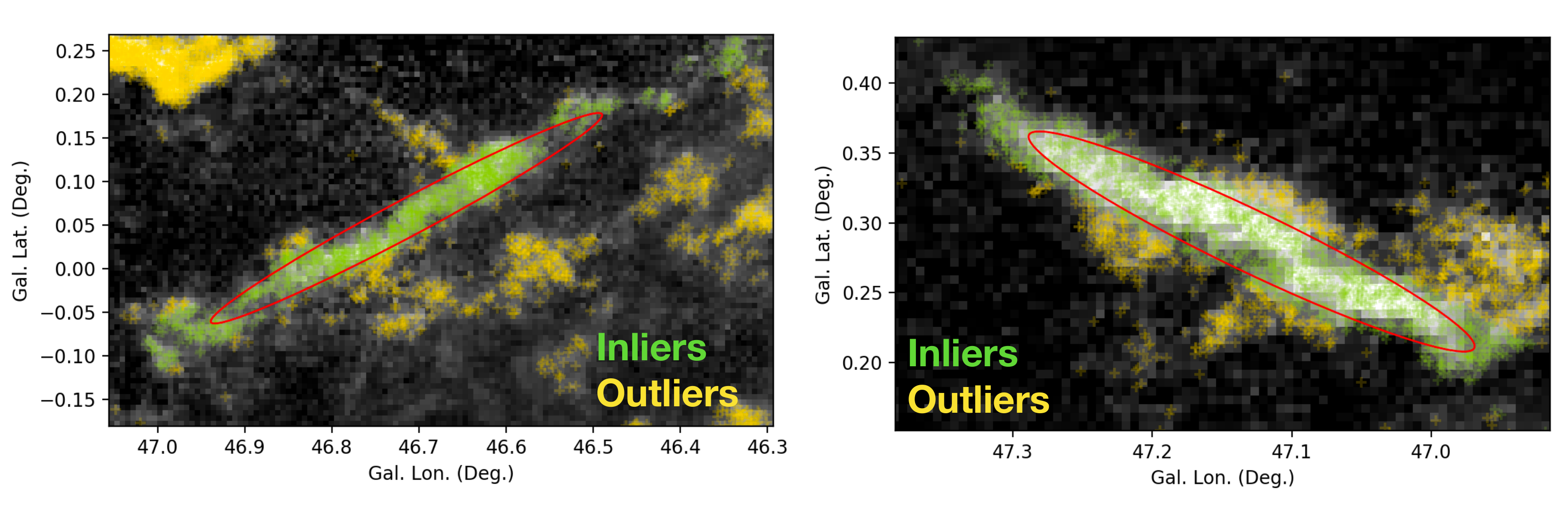}

  \caption{ Contrail separation and measurement of aspect ratios. The left and right panel present results from the G46 contrail and the G47 contrail, where the grayscale image represent the velocity-integrated $^{13}$CO(1-0) emission. The green crosses mark the region considered to be inside the contrails where as the yellow crosses mark the region that does not belong to the contrails. The red ellipses are obtained by diagonalising the tensor
    of second moments of positions of points that are associated with the contrails. From these fittings, we estimate an aspect ratio of 10 for the G46 contrail and an aspect ratio of 6 for the G47 contrail.  \label{fig:fit}
    % Right panel \label{fig:fit}, the points represent the spatial distribution of dense cores in the W43 region. The red points are cores that belongs to the contrail. The pink ellipse is obtained by fitting an ellipse to points that are associated with the contrails.
  }

\end{figure*}

{
  To refine the estimation of  the aspect ratios of the contrails, we adapt the following two-step
  approach: first, we make the use of the \texttt{ransac (RANdom SAmple
    Consensus)} algorithm \citep{fischler1981} to separate the contrails. The \texttt{ransac} algorithm is an iterative method
  capable of fittings models to data where a significant number of outliers
  are present in a robust fashion. In our case, the model is chosen to be a straight line, and by fitting it to our data we separate the contrails. After that, we fit
  ellipses to the points found to be associated with the contrails by diagonalising the tensor
  of second moments of positions of points that are associated with the contrails, and the
  aspect ratios of contrails are taken as the ratios between the long and short
  axes of the ellipses. The values we obtained through fitting do not differ significantly from the values we measured by hand.
  The results are presented in \ref{fig:fit}.
  We note that \texttt{ransac} is an algorithm that works on a set of data points. Therefore,
  to begin with, we have used the rejection sampling  method \footnote{\url{https://en.wikipedia.org/wiki/Rejection_sampling}}
  to convert our images into points upon which the \texttt{ransac} algorithm is applied. The \texttt{ransac} algorithm is not capable of dealing with cases where the outliers outnumbered the inliers. Therefore, we cropped our images in the position-position-velocity space such that the majority of the emission originates from the contrails.}

% Don't change these lines
\bsp	% typesetting comment
\label{lastpage}
\end{document}